\documentclass[12pt,preprint]{aastex}

\shorttitle{Deuterated Thioformaldehyde in B1}
\shortauthors{Marcelino et al.}

\begin{document}

\title{DEUTERATED THIOFORMALDEHYDE IN THE BARNARD 1 CLOUD}

\author{N. Marcelino}
\affil{IRAM, Av. Divina Pastora, 7, 18012 Granada, Spain}
\email{marcelino@iram.es}
\author{J. Cernicharo}
\affil{DAMIR, IEM-CSIC, 28006 Madrid, Spain}
\email{cerni@damir.iem.csic.es}
\author{E. Roueff}
\affil{LUTH (CNRS/UMR8102 and Observatoire de Paris), place J. Janssen, 92195
 Meudon cedex France}
\email{Evelyne.Roueff@obspm.fr}
\author{M. Gerin}
\affil{LERMA (CNRS/UMR8112 - Observatoire de Paris and Ecole Normale
 Superieure), 24 Rue Lhomond, 75231 Paris cedex 05, France}
\email{gerin@lra.ens.fr}
\and
\author{R. Mauersberger}
\affil{IRAM, Av. Divina Pastora, 7, 18012 Granada, Spain}
\email{mauers@iram.es}

\begin{abstract}
We present observations of the singly and doubly deuterated species of
thioformaldehyde, HDCS and D$_2$CS, towards the dark cloud Barnard 1. This is 
the first detection of D$_2$CS in Space and in dense and cold prestellar 
regions. Column densities obtained using rotational diagrams and a Large 
Velocity Gradient model show an extremely high D-enhancement in thioformaldehyde 
in Barnard 1. Although the column density of H$_2$CS is smaller than that of 
H$_2$CO, both species show similar D-enhancements in their singly and doubly 
deuterated species. A chemical model -including multiply deuterated species-
has been used in order to interpret the observations.

Predicted rotational frequencies from laboratory data for HDCS and
D$_2$CS are significantly in error when compared to the observed
frequencies in Space. Consequently, we have derived new rotational
constants for these two species and for H$_2$CS and H$_2$C$^{34}$S
using the observed frequencies in Barnard 1. The new rotational constants
allow to predict the rotational transitions of these species with
the accuracy needed for the narrow line emerging from dark clouds.
Rotational constants for HDCS and D$_2$CS have been obtained from the
observed transitions in the laboratory and in Space.

\end{abstract}


\section{INTRODUCTION}

For a long time, singly deuterated interstellar molecules are
known to be
present in
cold molecular clouds with an extremely high abundance enhancement due
to chemical fractionation \citep{van95,Min97,Sai00,Tin00,Tur01,Par02}.
The formation of deuterated molecules is preferred at low temperatures
($\leq$80 K) and leads to a high degree of fractionation in cold and
dense dark clouds without embedded luminous sources. These prestellar
cores offer the opportunity to study the cold and dense evolutionary
stage in which the deuteration of molecules supposedly occurred.
Although this is a relatively well understood chemical process,
the observation in the last few years
of multiply deuterated molecules
has renewed interest in deuterium fractionation in dense
clouds :
ND$_2$H \citep{Rou00}; ND$_2$H and D$_2$CO \citep{Loi01};
ND$_3$ \citep{Lis02}; D$_2$S \citep{Vas03}; D$_2$CO
\citep{Cec02,Bac03}; CHD$_2$OH and CD$_3$OH \citep{Par02,Par04}.
Among these deuterium isotopic species,
there are, so far, only three sulfur bearing
deuterated molecules,
HDS and D$_2$S, detected towards several protostars and dense cores
 \citep{Vas03,van95};
and HDCS, which has been detected toward TMC-1 \citep{Min97}.

Our new data and, in particular the high enhancement in the
fractionation of multiply deuterated species, are requiring a review of
chemical models.
The presence of deuterated molecular species in star forming regions
like Orion-IRc2 and IRAS 16293-2422 \citep{Tur90,van95,Cec98,Loi00} has been
 taken
as signs for an active grain surface chemistry followed by
desorption when the physical processes associated to
recent star formation activity
increases the temperature of the gas and of dust grains.
However, this is not the case for dark clouds, where the
kinetic temperature is too low for efficient evaporation of ice grain
mantles.
\cite{Rob00} found that gas phase chemistry can account
for the large abundances of both singly and doubly deuterated
molecules when the gas is very depleted of molecules and atoms.
The reason is that both CO and O efficiently destroy H$_3^+$. Hence when CO
and O are sufficiently depleted, H$_3^+$ is converted efficiently into
H$_2$D$^+$ in the reaction with HD. Subsequent reactions of H$_2$D$^+$ with HD
leads to D$_2$H$^+$ and finally to the formation of D$_3^+$. These ions
propagate the deuterium to other molecules \citep{Lis02,Rob03}.
In their model, \cite{Rob03}, conclude that in regions where CO is
higly depleted, the abundance of D$_2$H$^+$ will be simillar than that
of H$_2$D$^+$ and that D$_3^+$ could be the most abundant form of
deuterated H$_3^+$. The first prediction has been confirmed recently
with the detection of D$_2$H$^+$ toward the prestellar
core 16293E \citep{Vas04}.
For an early evolution time ($\sim$2-4$\times$10$^4$ yr)
the double and triple deuterated species could have large enhancements
\citep{Rob03}.
In the case of Barnard 1, \cite{Lis02}, based on their C$^{18}$O and C$^{17}$O
(2-1) observations found that C$^{18}$O/H$_2$=5.4$\times$10$^{-8}$, which is
$\sim$3 times lower than the canonical value,
 C$^{18}$O/H$_2$=1.7$\times$10$^{-7}$
\citep{Fre82, Cer87}, indicating that
an appreciable
amount of CO is depleted onto grains.
Modelling with low temperature conditions and high depletion factors allows a
satisfactory comparison with the observations of doubly and triply deuterated
ammonia -see e.g. NHD$_2$ towards L134N \citep{Rou00}- and also helps to explain
 the high
abundances observed in star forming regions in terms of grain
chemistry models.
The recent detection of doubly deuterated H$_2$S provides a stimulating impetus
to consider other sulfur molecules and extend the previously elaborated
chemical schemes.

During a 3mm line survey on dark clouds (Barnard 1, TMC-1, L1544, L183
-Marcelino et al., in prep.), we
found several transitions that initially were considered as
unidentified. Taking into account the reduced number of possible
candidates, the fact that the lines were strongest in Barnard 1 (see Figure 1),
 and the
strong deuterium enhancement found in these sources by several
authors \citep[see, e.g.,][and references therein]{Rou00,Ger01},
we concluded that these lines were arising from D$_2$CS and HDCS.
While single deuterated thioformaldehyde had been detected at lower
frequencies in TMC-1 \citep{Min97};
this is now the first detection of D$_2$CS in Space.
 Once the carriers were identified,
and given the extremely large deuterium enhacement indicated by the
observations,
we concentrated our efforts and available observing time in the
source showing the strongest lines, i.e., Barnard 1. In Section 2 we present
the observing procedures and in Section 3 the main results. In this
section we discuss the rotational constants of thioformaldehyde and
its isotopomers. We also analyze the physical and chemical
conditions of Barnard 1 and derive the column densities for the different
thioformaldehyde isotopic species. Section 4 is devoted to the discussion
and to the chemical analysis of the results.

\section{OBSERVATIONS}
The observations were carried out using the IRAM 30-m radio telescope at
Pico Veleta
(near Granada, Spain) between 2003 October and 2004 February.
Spectral lines arising from singly and doubly deuterated
thioformaldehyde between 83 and 155 GHz were observed in October.
In January we performed most of the observations for the main isotopic
species.
Up to four SIS receivers were used simultaneously. Each receiver was
tuned to single side band with image
rejections within  20-27 dB (3mm receivers), 12-16 dB (2mm receivers)
and 13 dB (1mm receivers).
System temperatures were
100-225 K in October for the 3 and 2mm receivers with
atmospheric opacities between 0.05-0.1. In 2004 January weather
conditions were even better allowing excellent observations at 1mm.
During this run we obtained temperatures below 100 K at 3mm and
below 200 K at 1mm (atmospheric opacities at 225 GHz were below 0.1).

Calibration was performed using two absorbers at
different temperatures and the ATM code developed by \cite{Cer85}
and recently updated by \cite{Par01}.
An autocorrelator providing several backends with 40
kHz and 80 kHz of resolution was used. The velocity resolution was
between 0.08-0.23 kms$^{-1}$ depending on the frequency.
Telescope pointing and focus were checked every 1 and
3 hours respectively
on strong and nearby continuum sources. The largest pointing error
between consecutive pointing scans never exceeded 3".

Most of the observations were performed using wobbler and frequency
switching techniques, using for the latter a frequency throw of 7.2
MHz -which eliminates the main ripple component coming from the telescope-
and in some cases we used position switching. The throw for
wobbler and position switching was 240" and 10' respectively,
both in azimut. Taking into account the size of the source, these offsets
are enough to provide a clean reference. On-source integration times were
 typically $\sim$1-2h,
resulting in rms main beam antenna temperatures between 15-50 mK.
Observed line parameters obtained from Gaussian fits to the observed lines
are given in Table 1 and the line profiles are shown in Fig. 2-4.
Line intensities for the observed transitions are given in
main beam brightness temperature scale. The main beam telescope
efficiencies are 0.78-0.75, 0.69 and 0.57 at 3, 2 and 1mm respectively.

In order to compare the deuterium enhancement in thioformaldehyde with
that of formaldehyde we also observed several rotational transitions of
H$_2$CO, HDCO, D$_2$CO and H$_2^{13}$CO in 2004 February.
The observed line parameters are
given in Table 2 and the corresponding line profiles are shown in Figure 5.

\section{RESULTS}
Figure 2a shows the observed lines for the singly and doubly
deuterated species of thioformaldehyde. The fact that the same
transitions of both molecules appear at different velocities when
assuming the predicted
frequencies from available line frequency catalogs -see, e.g., JPL
\citep{Pic98}; K\"oln \citep{Mul01}; Lovas \citep{Lov92} and Cernicharo
\citep{Cer00}- indicates a considerable error in the frequencies used.
We have used the new frequencies for these species derived from our observations
to improve the molecular constants and to obtain better values for
line prediction at higher frequencies (see Section 3.1). Figure 2b shows
that the agreement between HDCS
and D$_2$CS line profiles is excellent when the new predicted frequencies
for both species are used.

In Figure 2 we can see that lines arising from HDCS are moderately
stronger than those from D$_2$CS, which, although indicating a
large column density
for the singly deuterated species also implies a very high D-enhancement
for D$_2$CS (see below).
Line widths are rather narrow as
indicated in Table 1.
The similarity between line profiles from both species indicates
that they coexist spatially.

We observe a double peak profile for H$_2$CO lines probably due to
selfabsorption by foreground gas (the high velocity gas associated
with IRAS 03301+3057, a young stellar
object which is close to the strong ammonia peak, could also, perhaps,
affect
the line profile of formaldehyde; we have checked that the absorption feature
is not coming from emission in the off position by checking the
spectra in frequency switching mode).
Lines from the deuterated species of H$_2$CO and from H$_2$$^{13}$CO are
wider than the lines arising from
thioformaldehyde and its isotopomers, but less than those of H$_2$CO
and they do not show the line wings
observed in formaldehyde.
Lines arising from H$_2$CO are stronger than those from H$_2$CS and
lines from HDCO and D$_2$CO are about two times stronger than the
lines arising from deuterated thioformaldehyde, indicating a higher
column density for the different H$_2$CO isotopic species.

\subsection{Molecular Line Frequencies}
In order to have a coherent picture of the line profile from the
different transitions of thioformaldehyde  and its isotopes
we performed a determination of the velocity of the source as a
mandatory step in deriving line frequencies. We used 19
molecular lines that were observed in Barnard 1 during the present work,
and for which laboratory frequencies or confident frequency predictions
are available (JPL, K\"oln, Lovas, and/or Cernicharo
catalogs; see above). The lines showing some evidences of opacity effects were
discarded. Finally, only lines from CCH (some hyperfine components),
methanol, deuterated formaldehyde, CCS, and one of the hyperfine components
of the $N=1-0$ transition of CN were used. The averaged velocity obtained
from these lines is 6.70$\pm$0.05 km s$^{-1}$, where the error translates
the uncertainty from laboratory measurements and also the presence of a
second velocity component in the cloud at 7.5 km s$^{-1}$ that reduces
the accuracy of the Gaussian fits.
The estimated LSR velocity of 6.70 km s$^{-1}$ was assumed for all lines in
Table 1 when deriving observed frequencies.

We have observed four HDCS transitions,
based on the frequencies
calculated from the rotational constants of \cite{Min97}
and seven from D$_2$CS based on the frequencies derived
from the rotational constants of \cite{Cox82}.
The rotational
constants for the latter species have been derived from high $J$ transitions.
 However,
the largest frequency covered by the laboratory data is 35 GHz, hence,
one could expect significant uncertainties for rotational transitions at
large frequencies.
A comparison of the line profiles of HDCS and D$_2$CS (Fig. 2)
suggests frequency errors in the latter species as large as 0.1-0.2 MHz.
Even for HDCS for which the rotational constants have been derived from a
significant set of data \citep{Min97}, observed frequencies in dark clouds are
 often
different from those obtained in  the laboratory. As an example, the
3$_{03}$-2$_{02}$ line of HDCS has been observed by \cite{Min97} with a
laboratory frequency of 92981.601$\pm$0.018 MHz, while the same authors
reported a frequency for this line from observations of TMC-1
of 92981.680 MHz. We have observed the same line in Barnard 1 with much better
signal to noise ratio and we obtain (see Table 1) a frequency of
92981.658$\pm$0.030 MHz. Consequently, the use of the laboratory frequency
will produce  a 0.17 km s$^{-1}$  velocity
shift in the line -see Figure 2. It seems evident that
the calculated
frequencies in
the available line catalogs have significant errors for
some molecules. This is a critical
point, because astronomers often compare line profiles from different
 isotopomers
of a given molecular species to trace cloud collapse or line
opacities. Uncertainties in computed frequencies larger than 20 kHz
will produce significant velocity shifts when comparing different lines from
the same species or from different molecules.

The situation is even worse for
rare isotopic  substitutions like deuterium, $^{34}$S or $^{13}$C.
The rare isotopomer H$_2$C$^{34}$S is a much better tracer of the column
density of thioformaldehyde than the lines from the main isotope. We observed
 several
lines of this species (see Table 1) using frequencies from the JPL line
catalog. The number of measured laboratory frequencies for this species is
small and, hence, computed frequencies have considerable uncertainties.
For example
the 3$_{03}$-2$_{02}$ line predicted at 101283.899 MHz
was observed at 101284.331 MHz, the
3$_{12}$-2$_{11}$ line predicted at 102810.609 MHz was observed at
102807.426 MHz, the
4$_{14}$-3$_{13}$ predicted at 133031.473 MHz was observed at 133027.035 MHz,
and the 4$_{13}$-3$_{12}$ line predicted at 137077.543 MHz was observed at
137071.099 MHz. That the observed frequencies were corresponding to the
searched lines of H$_2$C$^{34}$S was confirmed by a simultaneous fit (see below)
 to
the laboratory and space frequencies of a few rotational constants (several
of them fixed to the values of the main isotope -note that H$_2$CS and
H$_2$C$^{34}$S have the same inertia moment around the A axis of the molecule).
The large differences that are found between space spectroscopy and
predicted frequencies could prompt for negative results and/or to reach
a confuse situation as we often look the on-line
frequency catalogs as a final product ready to be used. Special care has to
be taken
when using frequencies predicted from a fit to a reduced number of observed
lines, or for predictions in frequency domains outside those observed in the
laboratories.

Using the astronomically measured frequencies given in Table 1 and those
 observed in the
laboratories we performed fits to the rotational constants of H$_2$CS,
H$_2$C$^{34}$S, D$_2$CS and HDCS. The results are given in Tables 3 to 6. These
 tables also give
the fits obtained from the laboratory data alone. Although HDCS is a nearly
prolate symmetric top and a S-reduced Hamiltonian should be preferred
\citep{Gor84,Wat77,Min97}, we adopted the asymmetric (A) reduced form
of the Watson Hamiltonian in a I$^r$ representation \citep{Wat67,Wat77}.
Laboratory frequencies for the different isotopic species have
been taken from \cite{Clo83}, \cite{Dan78}, \cite{Chu73}, \cite{Cox82} and
\cite{Nau93}. Although it seems that the rotational constants
in Tables 3-6 show little variation between those obtained from
laboratory data and those derived from laboratory plus space observations,
these variations account for changes in the predicted frequencies of 0.1-0.2
MHz between both sets of data. The species the less affected by the
introduction of the space frequencies is HDCS and the most H$_2$C$^{34}$S.

One could think that the situation should be better for the main isotope of
thioformaldehyde for which 60 of its rotational lines have been measured
in the laboratory. However, Figure 3 shows that the laboratory frequencies
and the calculated ones result in different velocities for the lines.
Although, the uncertainties in the laboratory measurements are large, we
could expect that the global fit to all laboratory data will produce
rotational constants accurate enough to allow a good estimate of
the spectrum of H$_2$CS. As can be seen in the
central panels of Figure 3 this is really the case, but these predictions
are not accurate enough for the narrow lines emerging from dark clouds.
Hence, the frequencies of the eight observed lines of H$_2$CS were obtained
following a similar procedure to that adopted for its rare isotopes and a new
set of rotational constants was obtained. The results are given in Table 3.
With the new set of rotational constants for the four isotopomers of
thioformaldehyde, we have computed their frequencies and those corresponding
to the observed lines are given in Table 1.

The observed lines of H$_2$CO
show an asymmetric profile with a red line wing which could arise
from molecular outflows (see below).
For H$_2$CO, HDCO and also HDCS, the observed lines present a second peak
at $\sim$7.5 km s$^{-1}$. This velocity component is also observed
in HC$^{17}$O$^+$, CCH, c-C$_3$H and HC$_3$N.
\cite{Bac84} already found, based on the asymmetric and wide
$^{13}$CO line profiles, two velocity components in the cloud (at 6.5
and 7.2 km s$^ {-1}$).
The lines
of D$_2$CO peak at 6.7 km s$^{-1}$ (see Figure 5), while those of HDCO and
H$_2^{13}$CO have their emission maxima at slightly lower velocities.
If we assume for these lines a cloud velocity of 6.7 kms$^{-1}$, then
the frequencies for the lines of these two isotopes
of H$_2$CO
will be 0.1 MHz above of the
predicted ones, which are much more accurate (large number of
measured lines in the laboratory) than those of the
isotopes of H$_2$CS. Hence the observed velocities
for HDCO and H$_2^{13}$CO could correspond to
optical depth effects or to chemical cloud structure.
In fact, the high intensity
of the HDCO lines and the asymmetric profiles in H$_2$CO indicate
high optical depths in the Barnard 1 core, or high foreground column
densities, in the rotational lines of this species, even in those of HDCO.

\subsection{Column Densities and Physical Conditions}
\subsubsection{Previous Studies of Barnard 1}
Barnard 1 is one of the highest column density molecular sources in the Perseus
complex with recent low-mass star formation activity, as indicated by several
young stellar objects -three of them without optical counterpart- detected by
IRAS.
\cite{Bac84} made the first study of this region
using visual extinction maps derived from star counts \cite[see][]{Cer84},
and emission maps of
CO and its isotopomers, HCO$^+$, H$^{13}$CO$^+$
and NH$_3$. From star counts, they found Barnard 1 to be composed of two
condensations of A$_v$$\geq$5 mag.
They modeled the cloud in four regions -core, internal
and external envelope and halo- of different size and
physical conditions depending on the
line intensities of the observed molecules \cite[see also][]{Cer87}.
The innermost layer, i.e. the core, was studied from ammonia
data. With a visual extinction $\sim$10 and a size of 0.5 pc (at 200
pc of distance), the
average density in the core ranges from 10$^3$ to 3$\times$10$^4$ cm$^{-3}$ and
the kinetic temperature is 12$\pm2$ K (in concordance with CO data).
H$^{13}$CO$^+$ and HCO$^+$ emission is also detected in this region
with column densities N(H$^{13}$CO$^+$)$\sim$2$\times$10$^{12}$ cm$^{-2}$
and N(HCO$^+$)$\sim$2$\times$10$^{14}$ cm$^{-2}$, where the last one extends
to the internal envelope.
The rest of the cloud can be characterised by different isotopomers
of CO.
The density of the internal envelope was obtained from C$^{18}$O
emission and results to be 2$\times$10$^3$ cm$^{-3}$.
This region is very extended ($\sim$1 pc) and contains the young
stellar objects and the NH$_3$ globule. The visual extinction here is $\sim$3-4
 mag.
The external envelope, with a density below 10$^3$ cm$^{-3}$ and
A$_v$$\sim$2-3 mag, was
studied from $^{13}$CO emission.
The halo (A$_v$$\sim$1-2 mag) is the most external zone with a large spatial
 size
and traced by moderate/weak CO emission -see also \cite{Cer87}.
They also made a dynamical study of the cloud and concluded that it is
being supported by magnetic fields and/or turbulence.

\cite{Bac90} mapped Barnard 1 in several molecules with higher angular
resolution, resolving some spatial structures in the main dense core from
observations of NH$_3$. In this region they found density values of
7$\times$10$^4$ cm$^{-3}$ in the southern clump, and $>$3.7$\times$10$^4$
 cm$^{-3}$
in the northern one.
They studied the external and the internal envelopes from $^{13}$CO,
C$^{18}$O and CS emission. CS maps show the dense region to have a
size of $\sim$ 2 pc $\times$ 5 pc (at 350 pc of distance).
The strongest
emission of this molecule is restricted to the internal envelope, but
extends to the external one with lower intensities, probably due to
the existence of several dense clumps.
They obtained $n$(H$_2$)$>$2$\times$10$^3$ and 10$^3$ cm$^{-3}$ for the internal
and the external envelopes, respectively.
The kinetic temperature is 11-14 K in the densest regions and
increases to T$_{\rm kin}>$20 K at the edge of the cloud, showing that the
cloud is submitted to some heating from the exterior.

Continuum observations indicate the presence of four clumps in Barnard 1
\citep[see][]{Mat02}. Among them, B1-b corresponds to the position we are
observing and the source IRAS03301+3057 is close to B1-a. Barnard1-b has two
intense (sub)millimeter continuum sources, probably Class 0 sources associated
to compact protostellar objects \citep{Hir99}. However, no outflows have been
detected from these objects. Based on their dust continuum observations,
\cite{Hir99} obtained a H$_2$ column density $\sim$1.3$\times$10$^{23}$
 cm$^{-2}$ in a
$15''$ beam. Assuming a distance of 200 pc, this column density corresponds
to a mean gas density of 2.9$\times$10$^6$ cm$^{-3}$ in the beam diameter,
 showing that
high gas densities are present in the source.
From CO $J=2-1$ observations, \cite{Bac90} found a
high velocity flow emerging from IRAS 03301+3057, a young stellar object located
near the strong ammonia peak (B1-a). This outflow, which covers a region of
about $40''$, was studied in detail by \cite{Hir97} from CO ($J=1-0$)
emission and by \cite{Yam92} from SiO $J=2-1$ and $J=1-0$ lines.
The interaction of the flow with the ambient gas is traced by
the different Herbig-Haro objects present in this cloud
\citep{Yan98}.

The wings in the H$_2$CO lines shown in Figure 5 could be produced
by the high velocity gas associated to
IRAS 03301+3057 (located in B1-a). Note, however, that
our position (the B1-b clump) is about $1'$ East from this source,
and from the compact SiO peak which marks the place were the
outflow interacts with the cloud material. Hence, the wings could be
unrelated to the deuterium source.

\subsubsection{Molecular Background}
Formaldehyde and thioformaldehyde have both a nuclear spin statistics
that separates the rotational levels into two different sets. Radiative
and collisional transitions between ortho and para species are strictly
forbidden, hence, these two species behave like two different molecules.
Rotational levels
with k$_{-1}$ and k$_{+1}$  values even/odd or even/even belong to the para
species while those with odd/even and odd/odd belong to the ortho species.
For H$_2$CO, H$_2$CS, and their $^{13}$C isotopic species the nuclear
spin statistics leads to a 3:1 value for the ortho/para ratio.
The singly deuterated species HDCO and HDCS do not have this separation
on the rotational levels because the symmetry of the molecule has been broken
by the deuteration. However, these species have b-type rotational transitions
as deuteration introduces a weak permanent dipole moment along the B axis
of the molecule (typically 0.2-0.3 D). Doubly deuterated species on
the other hand also have
a symmetry axis, and the nuclear spin statistics separate the rotational
levels. However, as the nuclear spins are different, rotational levels
of doubly deuterated
isotopic species with k$_{-1}$ and k$_{+1}$ values odd/even or odd/odd
belong to
the para species, and those with even/odd even/even to the ortho one, i.e., just the opposite
than  for the main isotopic species. Moreover, the ortho/para ratio
in the doubly deuterated species of formaldehyde and thioformaldehyde is
2.

\subsubsection{Thioformaldehyde}
In order to derive column densities for thioformaldehyde and their isotopes
we have proceeded in two different ways.
The first method we have used consists in the
analysis of rotational diagrams for the observed isotopes of thioformaldehyde,
see, e.g., \cite{Gol99}. From these diagrams one can obtain the
molecular column
density and the excitation temperature of the molecular rotational levels
(assuming all levels are populated by a single excitation temperature).
From these diagrams we obtain common rotational temperatures close to 10 K,
$T_{\rm rot}$=9.5$\pm$1.0, 10.2$\pm$2.0, and 7.4$\pm$0.8 K for HDCS, D$_2$CS and
H$_2$CS respectively, which suggests that the molecular hydrogen volume
density is large enough to
thermalize the rotational levels of these molecules (in the case of
thermalization, $T_{\rm rot}$ should be close to $T_{\rm kin}$).
The derived
column densities for these species, assuming that the emitting source
fills the beam of the 30-m radio telescope, are N(H$_2$CS)=1.2-2.0$\times$10$^{13}$
cm$^{-2}$, N(HDCS)=4.5-7.0$\times$10$^{12}$ cm$^{-2}$,
N(D$_2$CS)=1.0-4.0$\times$10$^{12}$ cm$^{-2}$.
Hence, the abundance ratios are N(H$_2$CS)/N(HDCS)$\simeq$3$\pm$1,
N(HDCS)/N(D$_2$CS)$\simeq$3$\pm$2,
and N(H$_2$CS)/N(D$_2$CS)$\simeq$9$\pm$3.
Finally, for H$_2$C$^{34}$S we have assumed
the rotational temperature obtained for H$_2$CS to derive N(H$_2$C$^{34}$S)=
6.0$\pm$3.0$\times$10$^{11}$ cm$^{-2}$.
The column densities of H$_2$CS and H$_2$C$^{34}$S are compatible with
a galactic $^{32}$S/$^{34}$S ratio of 24.4$\pm$5.0 \citep{Chi96}.
This analysis does not provide any insight on the volume density
and assumes that the lines are optically thin (which seems to be correct
in view of the low column density derived for H$_2$C$^{34}$S).
The volume density and other physical parameters could be derived through
the observed line intensities and a Large Velocity Gradient code
\cite[see, e.g.,][]{Gol74} if the collisional rates are known.
We have adopted for H$_2$CS the collisional rates of formaldehyde
computed
by \cite{Gre78}, multiplied by a factor of 2 to take into account the
different effective surface and different energy of the levels of H$_2$CS
(a similar factor to that existing between the collisional rates of CO
and of CS).
Taking into account that we have observed a large
number of lines of
H$_2$CS, we could check that the adopted collisional cross sections produce
a good fit to the observed intensities using reasonable values for
the gas density and temperature. Of course, the final result has to be
in good agreement with the column densities and rotational temperatures
derived from the rotational diagrams. Assuming a kinetic temperature of
10 K (see section 3.2.1) we derive rotational temperatures for the observed
transitions of thioformaldehyde between 6.5 and 9.1 K for volume densities of
2-4$\times$10$^5$ cm$^{-3}$ and a column density of ortho and para species of
1.2 and 0.5$\times$10$^{13}$ cm$^{-2}$ respectively. These values are in reasonable
agreement with those derived from the rotational diagrams.

For D$_2$CS we have also used the same collision rates as for H$_2$CS.
Assuming also a kinetic temperature of 10 K, we
derive n(H$_2$)=3$\times$10$^{5}$ cm$^{-3}$, $T_{\rm rot}\simeq$10 K,
and N(D$_2$CS)$\simeq$3.0$\times$10$^{12}$ cm$^{-2}$,
with a derived value for the ortho/para ratio of $\simeq$1.7, very close to
the expected value of 2.

For HDCS the collisional rates could be rather different from those of
the main isotope. We have tried to fit the data using the
collisional rates for HNCO, but it was difficult to reach a good fit to
the data with
reasonable values for the physical parameters (which could be expected from
this crude approximation to the collisional rates of HDCS).
We adopt for HDCS
the values obtained above from the rotational diagrams (note that the LVG values
for the other species compare very well with those obtained from the rotational
diagrams).

\subsubsection{Formaldehyde}
The column density of formaldehyde and its isotopes have been derived
from an LVG analysis of the observed intensities, adopting the collisional rates
of \cite{Gre78}. The line profile of H$_2$CO indicates opacities larger than 1.
Hence, the column density for this species has been derived from the isotope
H$_2^{13}$CO assuming an $^{12}$C/$^{13}$C isotopic ratio for dark clouds of 50 
\citep{Cer87, Lan80}. Assuming again
a kinetic temperature of 10 K, the observed intensities give
N(H$_2^{13}$CO)=1.2$\pm$0.5$\times$10$^{12}$ cm$^{-2}$, n(H$_2$)=3$\times$10$^5$ cm$^{-3}$.
There is no indication for an ortho/para ratio different than 3 from the
observational data. The column density for the doubly deuterated species
is N(D$_2$CO)=3.0$\pm$1.5$\times$10$^{12}$ cm$^{-2}$. For the single deuterated
species we have assumed the same rotational temperatures as for the other
isotopes of formaldehyde to obtain N(HDCO)=7$\pm$2$\times$10$^{12}$ cm$^{-2}$.
Hence, the isotopic ratios are N(H$_2$CO)/N(HDCO)$\simeq$7,
N(H$_2$CO)/N(D$_2$CO)$\simeq$13, and N(HDCO)/N(D$_2$CO)$\simeq$2.5. While
the abundance ratios of singly over doubly deuterated species for formaldehyde
and thioformaldehyde are similar ($\sim$3), the abundance ratio between the
main isotope and the singly deuterated species is twice larger in thioformaldehyde
compared to formaldehyde.

\section{ABUNDANCE RATIOS, DEUTERIUM FRACTIONATION AND CHEMICAL MODELLING}
Table 7 sumarizes several studies in dark clouds and star forming
regions where deuterated molecules have been found. For Barnard 1, the new
ratios obtained in this work for H$_2$CS and H$_2$CO are also shown. Comparing
with other deuterated species detected towards this core, we can see that the
D-enhancement of thioformaldehyde is among the highest values. In general the
abundance ratios in Barnard 1 are only comparable with those found in the class
0 binary protostar IRAS 16293-2422. That means that deuteration processes should
be similar in both sources and different than in dark and quiescent clouds.

We have extended our chemical network dedicated to deuterium containing
molecules \cite[][Roueff et al. 2004, in prep.]{Lis02} to sulfur bearing species.
SD, HDS, D$_2$S, HDCS and D$_2$CS, SD$^+$, HDS$^+$, D$_2$S$^+$, H$_2$DS$^+$,
 HD$_2$S$^+$ ,D$_3$S$^+$, HDCS$^+$, D$_2$CS$^+$, H$_2$DCS$^+$, HD$_2$CS$^+$
and D$_3$CS$^+$ have been added to the previous chemical network.
We have assumed, as is usual in similar studies, that deuterated
species behave as the main isotopic species and we have calculated the
different branching ratios from statistical arguments except when
dissociative recombination reactions were considered. In this latter
case, we have assumed that removal of an hydrogen atom was more
efficient that the release of D by a factor of 2. This has indeed been
seen in various experimental studies of multiply deuterated molecular
ions as discussed by \cite{Pet03}. The chemical reaction
rate coefficients have been taken from our own chemical network
\citep{Pin93}, updated with the recent compilation of \cite{Ani03} and from
the available chemical network of the Ohio State University (Herbst and coworkers).
As is recalled in \cite{Mil03}, deuterium enrichment is initiated by the
fractionation reactions of H$_3^+$, CH$_3^+$ and C$_2$H$_2^+$
with D and HD.
We have extended the reactions to the partially deuterated species until the
fully deuterated ions D$_3^+$, CD$_3^+$ and C$_2$D$_2^+$ are formed
-as studied in the laboratory by \cite{Ger02b}.
Another important addition consists in considering the fractionation
reactions with D$_2$, which allows more substantial deuterium
enhancement. Recent laboratory studies have shown that the reactions
are significantly less efficient as currently assumed
\cite[cf.][]{Mil03} and that the reverse endothermic reaction may be more rapid than
expected from pure thermodynamics. We have thus taken the reaction
rate coefficients derived from the experiments \citep{Ger02a} but
determined the reverse reactions with the calculated endothermicity.
We have also introduced possible fractionation reactions of
H$_2$CS$^+$, H$_3$CS$^+$ and H$_3$S$^+$ with HD, which could occur but
have not been considered until now as no experimental or theoretical
information is available. The corresponding reactions and rate coefficients
are displayed in Table 8 where the values of the
endothermicities involved in the backward reactions have been guessed. As
the Langevin collision rates have been assumed, the results obtained
by including these reactions will give upper limits for the deuterium
fractionation.

Their influence is discussed below. We did not include the chemistry of methanol.
The resulting chemical network consists in 210 species linked by more
than 3000 chemical reactions. Details and discussion on the deuterated
chemistry are discussed in Roueff et al. (2004, in prep.) and we focus the present
discussion to the sulfur compound deuteration.
Recent observations have shown that enhanced deuteration is linked to large
depletions \cite[see for example][]{Bac03}. We consider then three different models
with H$_2$ densities of 10$^4$, 10$^5$ and 10$^6$ and different
elemental depletions as displayed in Table 9. These three
models correspond approximately to the conditions found in TMC-1, Barnard 1 and
L134N \cite[see,][Pagani et al. 2004, in press]{Cer87,Bac90}.
The cosmic ionization rate and the temperature are kept constant and
equal respectively to 2$\times$10$^{-17}$ s$^{-1}$ and 10 K.
The sulfur abundance is also taken constant and typical of the low metal case, i.e.
S/H$_2$ = 3.70$\times$10$^{-7}$.
The Table shows that, as expected, deuteration increases
from models 1 to 3. Fractionation reactions involving H$_2$CO$^+$,
H$_3$CO$^+$, H$_3$S$^+$ and H$_3$CS$^+$,
are seen to be of
considerable importance for the fractionation of H$_2$S. However,
their role is much less for formaldehyde and thioformaldehyde. This is
readily explained by the chemistry as  hydrogen sulfide is directly
obtained from dissociative recombination of H$_3$S$^+$ and involves a
succession of ion molecule reactions. In the case of formaldehyde and
thioformaldehyde, neutral reactions involving CH$_3$ and atomic oxygen
and atomic sulfur are supposed to occur at a rate of 1.4$\times$10$^{-10}$.
Then, deuterium enhancement of H$_2$CO and H$_2$CS
reflects directly the deuterium enhancement of CH$_3$. The reaction of
CH$_3$ and O has only been studied in the laboratory down to 300 K and
the reaction rate coefficient with sulfur is taken from the chemical
network of the Ohio State University in the group of E. Herbst.

As can be seen from the Table 9, this gas phase chemical
model gives the correct orders of magnitude of the deuterium fractionation
of ammonia, HCO$^+$, N$_2$H$^+$ for the high densities and large depletion
factors when comparing to the observed values. In this model grains play
a passive role, as they have trapped an important fraction of carbon and
oxygen, allowing the gas phase chemical processes to proceed. We also give the
fractional abundances of formaldehyde, sulfur hydride and thioformaldehyde to
check the relevance of our gas phase chemical network. We note that the
fractional abundance of sulfur hydride is between 10$^{-9}$ and
4$\times$10$^{-10}$ for the three models. This compares reasonably to the values
found in TMC-1 and L134N as reported by \cite{Vas03}. Grain surface chemistry is
often invoked for the formation of H$_2$S as the S$^+$ + H$_2$ reaction is
highly endothermic. In the present chemical network, SH$^+$ is formed via S +
H$_3^+$, which subsequently reacts with H$_2$ via radiative association to
produce H$_3$S$^+$. H$_2$S is then obtained from the dissociative recombination
of H$_3$S$^+$. The fractional abundance derived for H$_2$CS is below the
observed ones by values of a few. However, we have checked that increasing the
elemental abundance of sulfur results in increasing the steady state fractional
abundances of H$_2$S and H$_2$CS. We do not try to obtain a perfect fit with
the observed values as many physical and chemical parameters are not well
constrained. The present stage of agreement is nevertheless encouraging.

\section{CONCLUSIONS}
We have carried out a comprehensive observational study in Barnard 1 of the
deuterated species of thioformaldehyde, together with the observation of several
lines from the main isotope and from the $^{34}$S isotopic substituted species.
In order to compare the deuterium enhancement we have also observed formaldehyde
and its deuterated species. The main results are:

1) The systematic velocity of the dense gas in Barnard 1 has been determined to
be 6.70$\pm$0.05 km s$^{-1}$ from the observation of near 20 optically thin
molecular lines with well known frequencies.

2) The observed frequencies for HDCS, D$_2$CS and H$_2$C$^{34}$S as well as
the available laboratory data have been fit to obtain a set of molecular
rotational constants allowing precise frequency determinations for these
species in the millimeter domain.

3) Rotational temperatures and column densities for H$_2$CS, H$_2$CO and their
isotopes have been derived from rotational diagrams or through the use of a LVG
code. The derived H$_2$CS/HDCS and H$_2$CS/D$_2$CS abundance ratios are $\sim$3
and 9 respectively, while for formaldehyde are $\sim$7 and 13.

4) Steady state gas phase chemical models including multiply deuterated species
and with appropriate depletion factors, such as those found in different dense
cores, allow to reproduce the deuterium fractionation approximately. In this
context, surface processes have not to be included specifically , as they are
poorly known, and their role is essentially passive as the abundant oxygen, CO,
H$_2$O ...  components are frozen on the grains and are not available for gas
phase chemical processes to proceed. Our results involve also some specific
chemical assumptions which contribute to the high deuterium fractionation, namely
(1) specific branching ratios in dissociative recombination of
deuterated molecular ions with a 1:2 ratio for the ejection of H compared to D
as observed in several experiments. And
(2) fractionation reactions of
H$_2$CO$^+$, H$_3$CO$^+$, H$_3$S$^+$, H$_2$CS$^+$ with HD. These processes have
not been studied in the laboratory nor theoretically and such studies would be
very desirable. However, their influence on our results is not determinant
except for the deuterium fractionation of H$_2$S.

\acknowledgments
We would like to thank Spanish Ministry of Science and Technology, MCyT,
for funding support under grants ANAYA2000-1784, ANAYA2003-2785 and
PNIE2001-4516.
The authors are also grateful for detailed reading and useful suggestions of
the anonymous referee, which helped to improve this paper.

\clearpage

\begin{figure}
\includegraphics[angle=-90]{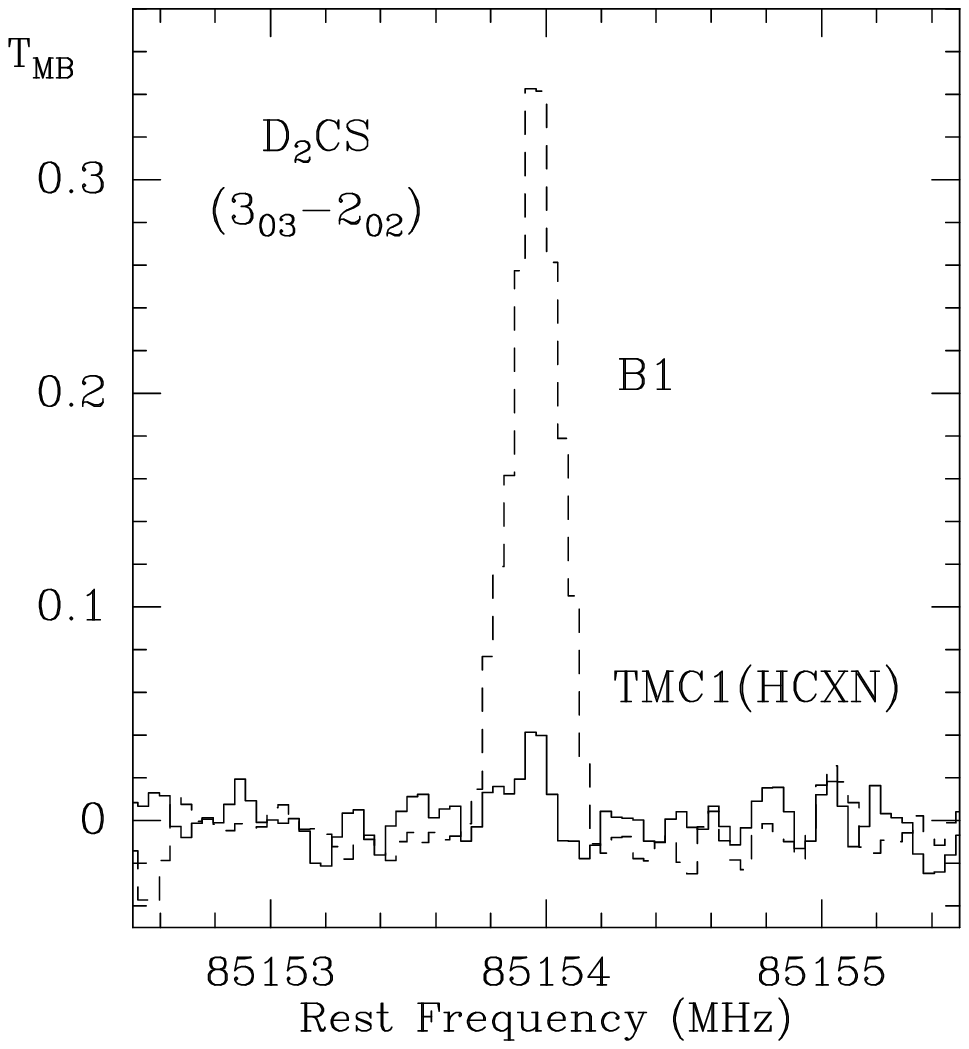}
\caption{D$_2$CS transition for Barnard 1 (dashed line)
and for TMC-1 (solid line). The position of TMC-1 is that of cyanopolyyne peak
RA(1950.0)$=04^{\rm h}38^{\rm m}38^{\rm s}.0$, dec(1950.0)$=+25^{\circ}35'45''$; that
 of Barnard 1
is RA(2000.0)$=03^{\rm h}33^{\rm m}20^{\rm s}.8$, dec(2000.0)$=+31^{\circ}07'34''$.
\label{fig1}}
\end{figure}

\clearpage

\begin{figure}
\plotone{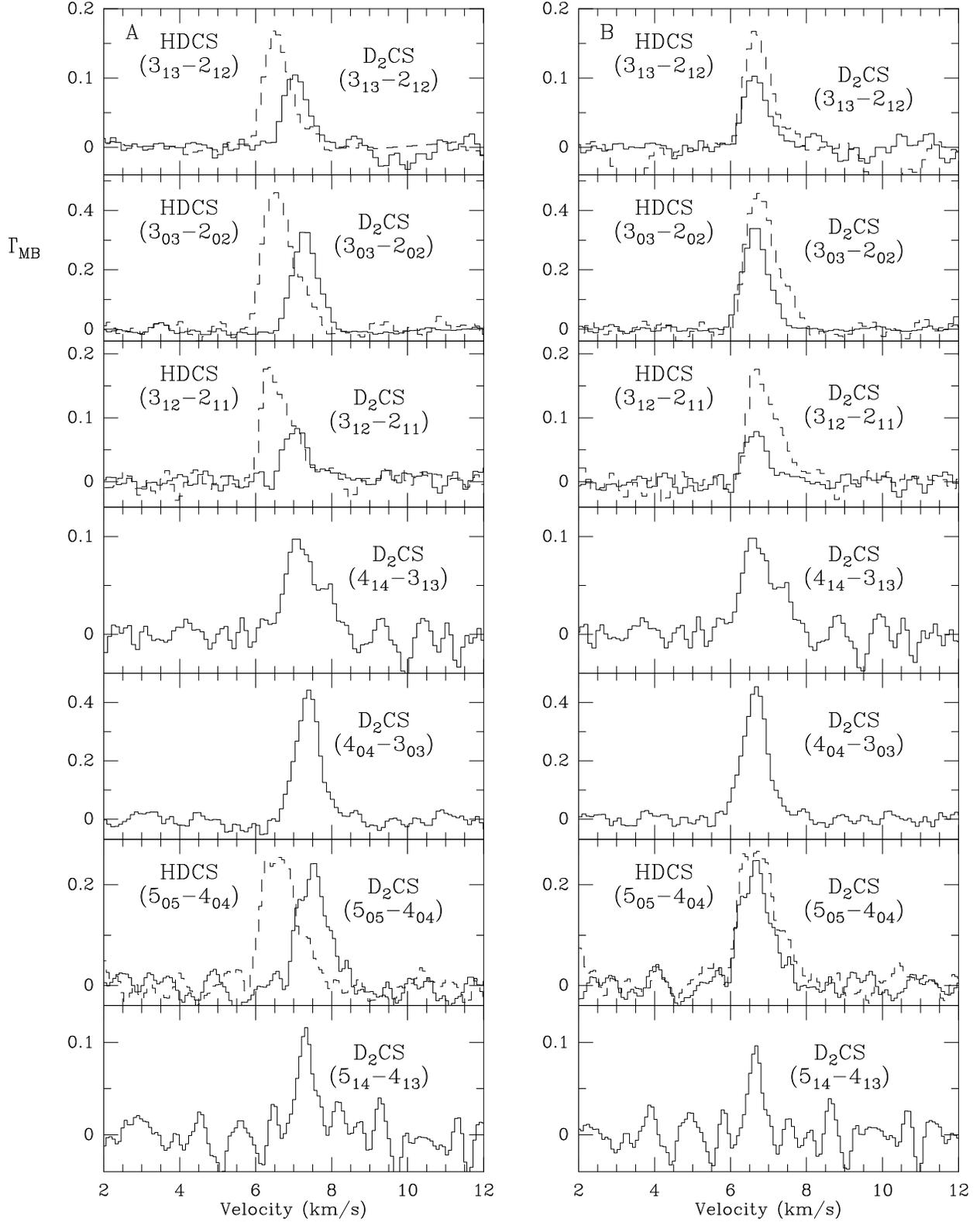}
\caption{Observed transitions for HDCS (dashed line) and D$_2$CS with
the frequencies originally adopted from the literature (a) and the
corrected frequencies (b).\label{fig2}}
\end{figure}

\clearpage

\begin{figure}
\epsscale{0.8}
\plotone{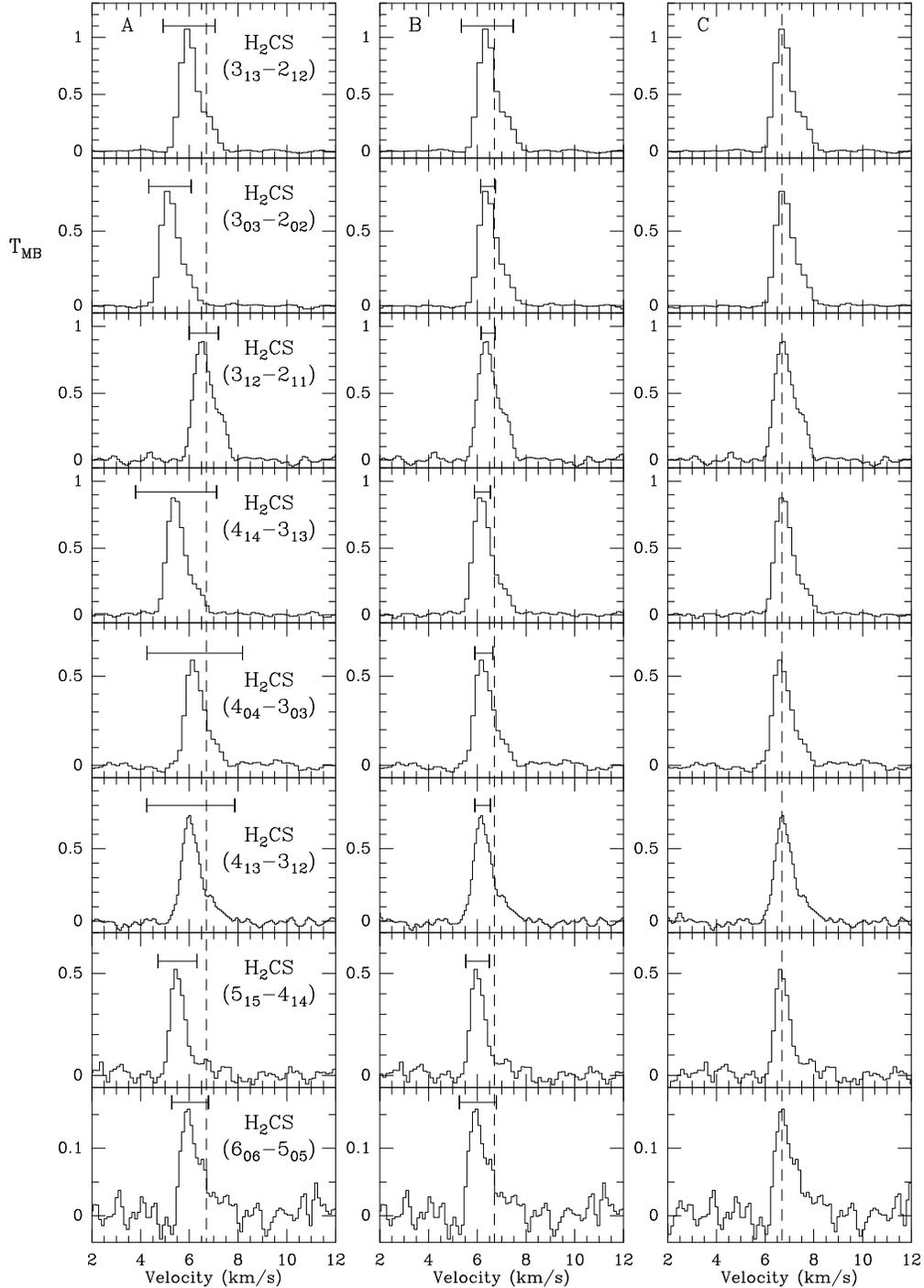}
\caption{Observed transitions for H$_2$CS. The velocity scale in the left
panels (A) has been derived from the laboratory frequencies, while for the
central panels (B) the predicted frequencies from a fit to the
laboratory data were used (see text). In the plots of these two columns
the horizontal lines represent the 3$\sigma$ uncertainty for
line frequencies (measured or predicted). Finally right panels (C)
show the H$_2$CS lines when a velocity of 6.7 km$^{-1}$ is assumed for
the main velocity component.\label{fig3}}
\end{figure}

\clearpage

\begin{figure}
\epsscale{0.6}
\plotone{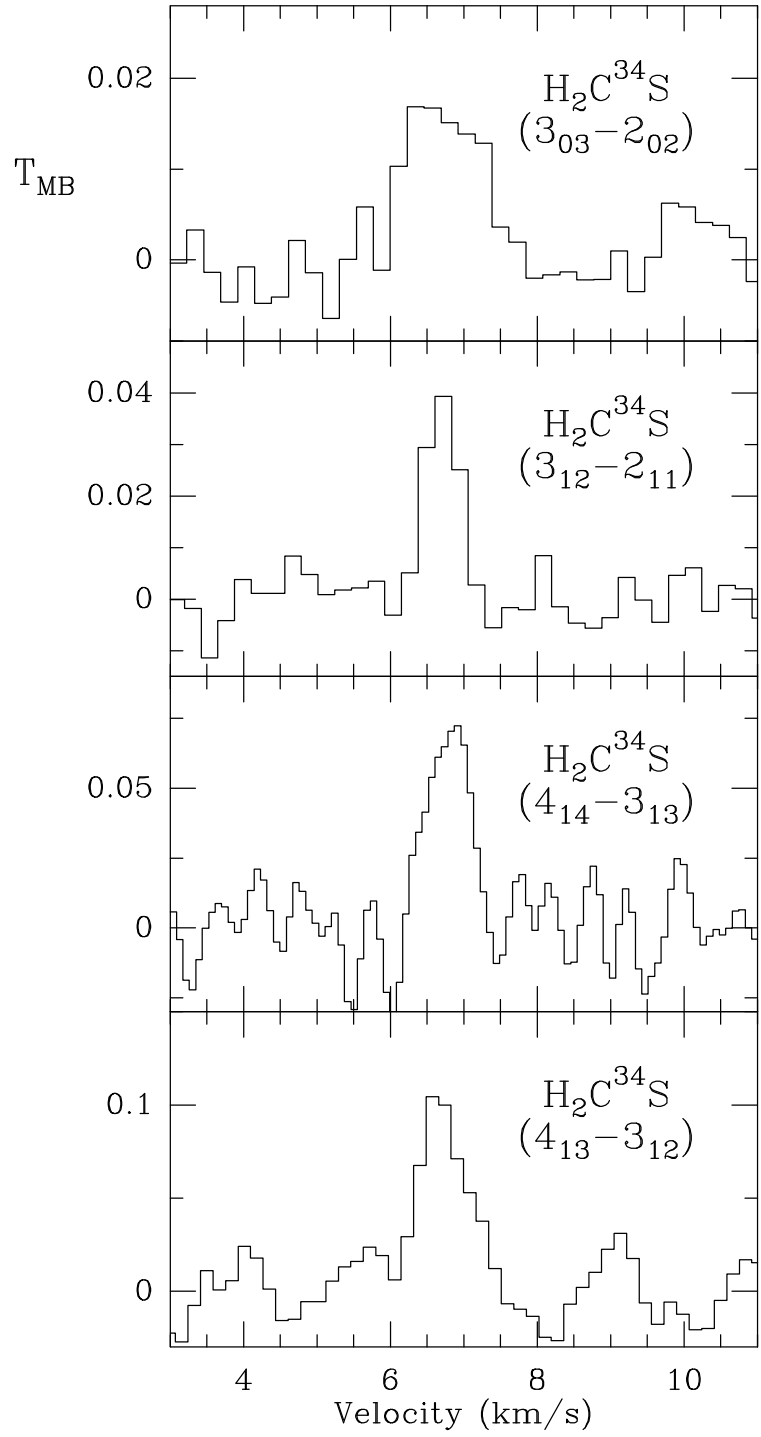}
\caption{Observed transitions of H$_2$C$^{34}$S in the direction of Barnard 1.
\label{fig4}}
\end{figure}

\clearpage

\begin{figure}
\epsscale{0.55}
\plotone{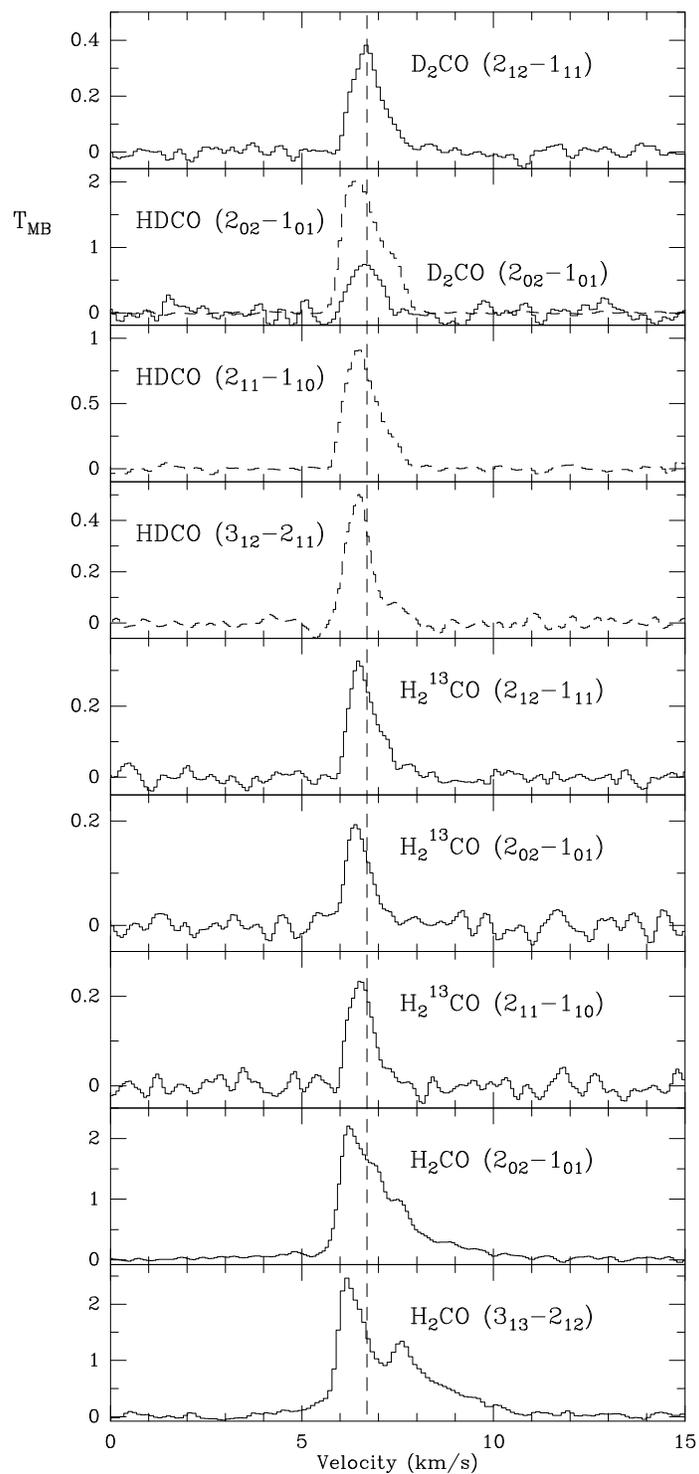}
\caption{Observed transitions from the different isotopic species of
formaldehyde. Velocity scale has been derived for the main isotope from
the calculated frequencies for this species, while for the rare species
the measured space frequency, assuming a velocity of 6.7 km$^{-1}$,
have been adopted (see text). \label{fig5}}
\end{figure}

\clearpage

\begin{figure}
\epsscale{0.65}
\plotone{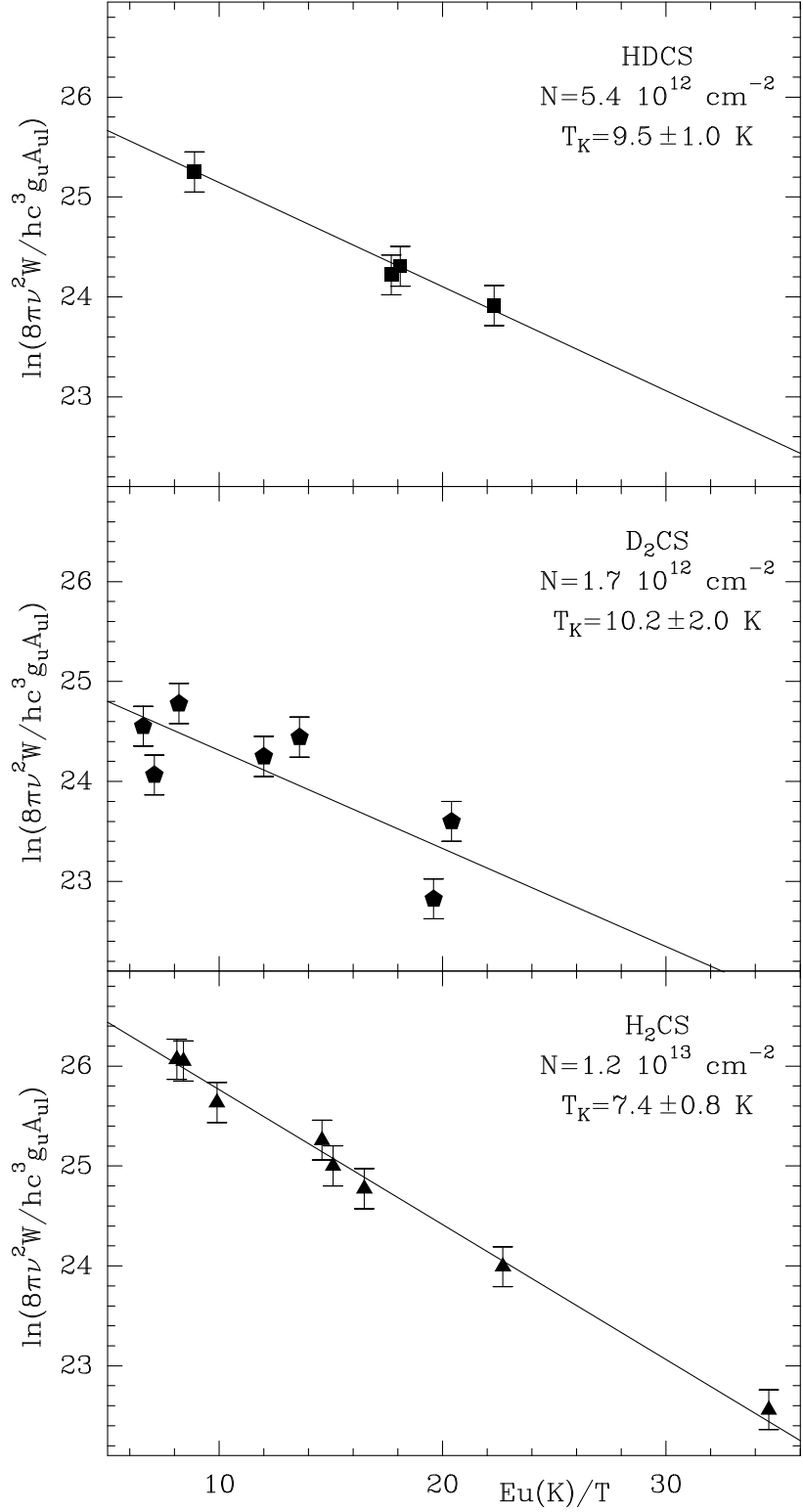}
\caption{Rotational diagrams for HDCS, D$_2$CS and H$_2$CS. The
errors are given between brackets and concerns the last significative
digit. Column density values are the averaged ones.  \label{fig6}}
\end{figure}

\clearpage

\begin{table}
\caption{Line parameters for the observed transitions of
thioformaldehyde species}
\vspace{0.2cm}
\begin{tabular}{c c r r r r}
\hline
\hline
         &            & \multicolumn{1}{c}{Observed}  & \multicolumn{1}{c}{Calculated} \\
Molecule & Transition & \multicolumn{1}{c}{Frequency\tablenotemark{a}} & 
\multicolumn{1}{c}{Frequency} & \multicolumn{1}{c}{$T_{\rm MB}$} &
\multicolumn{1}{c}{$\Delta v$}  \\
         &            & \multicolumn{1}{c}{(MHz)}                      & 
\multicolumn{1}{c}{(MHz)} & \multicolumn{1}{c}{(K)} & \multicolumn{1}{c}{(km/s)}   \\
\hline
HDCS & 3$_{13}$-2$_{12}$ &  91171.086(20)\tablenotemark{b}  &  91171.057 & 0.172( 10) & 0.760  \\
 & 3$_{03}$-2$_{02}$ &  92981.658(30) &  92981.607 & 0.467( 24) & 0.900  \\
 & 3$_{12}$-2$_{11}$ &  94828.592(41) &  94828.507 & 0.168( 19) & 0.884  \\
 & 5$_{05}$-4$_{04}$ & 154885.070(40) & 154885.053 & 0.278( 32) & 1.100  \\
D$_2$CS & 3$_{13}$-2$_{12}$ &  83077.759(24) & 83077.758 & 0.110( 14) &  0.754  \\
 & 3$_{03}$-2$_{02}$ &  85153.920(12) & 85153.932 & 0.339( 15) &  0.707  \\
 & 3$_{12}$-2$_{11}$ &  87302.662(24) & 87302.659 & 0.084( 12) &  0.638  \\
 & 4$_{14}$-3$_{13}$ & 110756.126(60) &110756.110 & 0.094( 19) &  1.222  \\
 & 4$_{04}$-3$_{03}$ & 113484.714(12) &113484.711 & 0.446( 35) &  0.682  \\
 & 5$_{05}$-4$_{04}$ & 141769.445(60) &141769.435 & 0.234( 23) &  0.874  \\
 & 5$_{14}$-4$_{13}$ & 145461.965(20) &145641.965 & 0.115( 29) & 0.402  \\
H$_2$CS & 3$_{13}$-2$_{12}$ & 101477.879( 6) &101477.878&1.054(  9)&  0.876\\
 & 3$_{03}$-2$_{02}$ & 103040.539(12) &103040.539& 0.766(  8)  &  0.891  \\
 & 3$_{12}$-2$_{11}$ & 104617.114(15) &104617.101& 0.856( 29) &  0.998  \\
 & 4$_{14}$-3$_{13}$ & 135298.407(15) &135298.373& 0.884( 13) &  0.870  \\
 & 4$_{04}$-3$_{03}$ & 137371.268(12) &137371.304& 0.586( 16) &  0.874  \\
 & 4$_{13}$-3$_{12}$ & 139483.741(24) &139483.706& 0.716( 34) &  0.855  \\
 & 5$_{15}$-4$_{14}$ & 169114.198(30) &169114.166& 0.541( 44) &  0.641  \\
 & 6$_{06}$-5$_{05}$ & 205987.954(60) &205987.908& 0.151( 24) &  0.832  \\
H$_2$C$^{34}$S & 3$_{03}$-2$_{02}$ & 101284.331(80)&101284.357& 0.018(  5)  &  1.170  \\
 & 3$_{12}$-2$_{11}$ & 102807.426(40)&102807.380& 0.042(  8)  &  0.497  \\
 & 4$_{14}$-3$_{13}$ & 133027.035(40)&133027.017& 0.078( 22) &  0.622  \\
 & 4$_{13}$-3$_{12}$ & 137071.099(30)&137071.093& 0.105( 16) &  0.758  \\

\end{tabular}
\tablenotetext{a}{
Observed frequencies have been estimated from a Gaussian fit
to the lines.
\\Quoted frequencies
correspond to the velocity component at 6.7 km s$^{-1}$}
\tablenotetext{b}{Errors for observed frequencies are given between
brackets}

\end{table}

\clearpage

\begin{table}
\caption{Line parameters for the observed transitions of formaldehyde species}
\vspace{0.2cm}
\begin{tabular}{c c r r l r}
\hline
\hline
Molecule & Transition & \multicolumn{1}{c}{Frequency\tablenotemark{a}} & 
\multicolumn{1}{c}{$v_{LSR}$} & \multicolumn{1}{c}{$T_{\rm MB}$} &
\multicolumn{1}{c}{$\Delta v$}  \\
         &            & \multicolumn{1}{c}{(MHz)}  &\multicolumn{1}{c}{(km s$^{-1}$)}\tablenotemark{b}  & \multicolumn{1}{c}{(K)} & \multicolumn{1}{c}{(km s$^{-1}$)}   \\
\hline
 HDCO & 2$_{02}$-1$_{01}$ & 128812.865 & 6.326 & 1.788( 17) & 0.689 \\
      & 		  &	       & 7.029 & 1.132      & 1.069 \\
      & 2$_{11}$-1$_{10}$ & 134284.909 & 6.389 & 0.781( 19) & 0.681  \\
      &			  &	       & 6.962 & 0.374      & 1.017 \\
      & 3$_{12}$-2$_{11}$ & 201341.377 & 6.459 & 0.516( 20) & 0.694 \\
      &			  & 	       & 7.454 & 0.080      & 0.613 \\

 D$_2$CO & 2$_{12}$-1$_{11}$ & 110837.821 & 6.272 & 0.164( 20) & 0.345 \\
	 &		     &		  & 6.698 & 0.354      & 0.567 \\
	 &		     &		  & 7.265 & 0.107      & 0.731 \\
         & 2$_{02}$-1$_{01}$ & 116688.442 & 6.658 & 0.817(121) & 0.839 \\

 H$_2$CO & 2$_{02}$-1$_{01}$\tablenotemark{c} & 145602.949 & 6.244 & 2.217( 42) & 0.590 \\
  	 &		     & 		  & 6.860 & 1.340 & 0.563 \\
	 &		     &		  & 7.513 & 0.722 & 0.795 \\
	 &		     & 		  & 8.483 & 0.264 & 2.264 \\
   & 3$_{13}$-2$_{12}$\tablenotemark{c} & 211211.448 & 6.095 & 1.348( 48) & 0.294 \\
   &			& 	          & 6.404 & 1.425 & 0.684 \\
   &			&		  & 7.592 & 0.769 & 3.215 \\
   &			&		  & 7.595 & 0.590 & 0.501 \\
 
 H$_2$$^{13}$CO & 2$_{12}$-1$_{11}$ & 137449.957 & 6.431 & 0.325( 24) & 0.480 \\
		&		    &		 & 6.802 & 0.128      & 0.423 \\
		&		    &		 & 7.156 & 0.100      & 0.420 \\
  & 2$_{02}$-1$_{01}$ & 141983.748 & 6.454 & 0.195( 20) & 0.733 \\
  & 2$_{11}$-1$_{10}$ & 146635.672 & 6.527 & 0.249( 22) & 0.681 \\
  &		      &		   & 7.315 & 0.040      & 0.206 \\	 

\end{tabular}
\tablenotetext{a}{Frequencies from Lovas catalog (see text)}
\tablenotetext{b}{Different line components}
\tablenotetext{c}{Complex line profiles with wings and selfabsorption}

\end{table}

\clearpage

\begin{table}
\caption{Rotational constants for H$_2$CS}
\vspace{0.2cm}
\begin{tabular}{l r@{.}l r@{.}l}
\hline
\hline
Parameter & \multicolumn{2}{c}{Laboratory} & \multicolumn{2}{c}{Lab+Sky} 
\\
\hline
A    & 291718&6 (133)          & 291708&3 (169)        \\
B    &  17699&6447 (146)        &  17699&6529 (168)      \\
C    &  16651&8711 (146)        &  16651&8587 (165)      \\
$\Delta_J$   &      2&0638 (391) $\times$ 10$^{-2}$     &      1&8916 (178) $\times$ 10$^{-2}$ \\
$\Delta_{JK}$  &    5&20998 (739) $\times$ 10$^{-1}$   &      5&22872 (573) $\times$ 10$^{-1}$ \\
$\Delta_K$   &      22&8519\tablenotemark{a}          &      22&8519\tablenotemark{a}      \\
$\delta_j$   &      1&21460 (129) $\times$ 10$^{-3}$  &      1&21438 (163) $\times$ 10$^{-3}$ \\
$\delta_k$   &      3&1922 (658) $\times$ 10$^{-1}$  &      3&2438 (836) $\times$ 10$^{-1}$  \\
H$_{J}$  &      1&056 (492) $\times$ 10$^{-5}$    &      -8&41 (313) $\times$ 10$^{-6}$    \\
\end{tabular}
\tablenotetext{a}{Value adopted from \cite{Clo83}}
\end{table}

\clearpage

\begin{table}
\caption{Rotational constants for D$_2$CS}
\vspace{0.2cm}
\begin{tabular}{l r@{.}l r@{.}l}
\hline
\hline
Parameter & \multicolumn{2}{c}{Laboratory} & \multicolumn{2}{c}{Lab+Sky} 
\\
\hline
A    & 146400&549 (284)         & 146397&685 (725)        \\
B    &  14904&7104 (116)        &  14904&70899 (556)      \\
C    &  13495&4298 (116)        &  13495&38776 (506)      \\
$\Delta_J$   &   1&2579\tablenotemark{a} $\times$ 10$^{-2}$     &  1&29708 (846) $\times$ 10$^{-2}$ \\
$\Delta_{JK}$  &      2&8869 (166) $\times$ 10$^{-1}$   &      2&7972 (178) $\times$ 10$^{-1}$ \\
$\Delta_K$   &      5&2084\tablenotemark{b}         &      6&588 (294)        \\
$\delta_j$   &      1&40587 (269) $\times$ 10$^{-3}$  &      1&38673 (369) $\times$ 10$^{-3}$ \\
$\delta_k$   &      2&14468 (275) $\times$ 10$^{-1}$  &      2&2554 (245) $\times$ 10$^{-1}$  \\
H$_{JK}$  &      1&198 (106) $\times$ 10$^{-6}$    &      7&25 (125) $\times$ 10$^{-7}$    \\
H$_{KJ}$  &     -3&6562 (919) $\times$ 10$^{-4}$   &     -3&075 (151) $\times$ 10$^{-4}$   \\
\end{tabular}
\tablenotetext{a}{Adopted value from  \cite{Cox82}}
\tablenotetext{b}{Force field value \citep{Tur81}}
\end{table}

\clearpage

\begin{table}
\caption{Rotational constants of HDCS}
\vspace{0.2cm}
\begin{tabular}{l r@{.}l r@{.}l}
\hline
\hline
Parameter & \multicolumn{2}{c}{Laboratory} & \multicolumn{2}{c}{Lab+Sky} \\
\hline
A            & 202716&66 (217)              &202716&54 (319)       \\
B            &  16111&38339 (610)           & 16111&38820 (887)    \\
C            &  14890&97584 (603)           & 14890&97943 (867)       \\
$\Delta_J$   &      1&587381 (547) $\times$ 10$^{-2}$ &     1&588513 (787) $\times$ 10$^{-2}$ \\
$\Delta_{JK}$&      3&180421 (533) $\times$ 10$^{-1}$ &     3&181521 (768) $\times$ 10$^{-1}$ \\
$\Delta_K$   &     14&72\tablenotemark{a}                &    14&72\tablenotemark{a}            \\
$\delta_j$   &      1&38002 (484) $\times$ 10$^{-3}$&     1&37782 (621) $\times$ 10$^{-3}$  \\
$\delta_k$   &      2&8831 (281) $\times$ 10$^{-1}$ &     2&8839 (414) $\times$ 10$^{-1}$   \\
H$_{JK}$     &      2&028 (217) $\times$ 10$^{-6}$  &     2&476 (312) $\times$ 10$^{-6}$    \\
H$_{KJ}$     &     -4&188 (109) $\times$ 10$^{-5}$  &    -4&193 (160) $\times$ 10$^{-5}$    \\
\end{tabular}
\tablenotetext{a}{Interpolated value from H$_2$CS and D$_2$CS}
\end{table}

\clearpage

\begin{table}
\caption{Rotational constants of H$_2$C$^{34}$S}
\vspace{0.2cm}
\begin{tabular}{l r@{.}l r@{.}l}
\hline
\hline
Parameter & \multicolumn{2}{c}{Laboratory} & \multicolumn{2}{c}{Lab+Sky} 
\\
\hline
A            & 291708&3\tablenotemark{a}                & 291708&3\tablenotemark{a}       \\
B            &  17389&1451 (120)           & 17389&1477 (108)    \\
C            &  16376&6721 (120)           & 16376&6745 (108)       \\
$\Delta_J$   &      1&8417 (486) $\times$ 10$^{-2}$ &     1&8283 (399) $\times$ 10$^{-2}$ \\
$\Delta_{JK}$&      5&0236 (169) $\times$ 10$^{-1}$ &     5&0328 (138) $\times$ 10$^{-1}$ \\
$\Delta_K$   &     22&8519\tablenotemark{a}                &    22&8519\tablenotemark{a}           \\
$\delta_j$   &      1&1662 (158) $\times$ 10$^{-3}$&     1&1673 (151) $\times$ 10$^{-3}$  \\
$\delta_k$   &      3&2438\tablenotemark{a} $\times$ 10$^{-1}$ &     3&2438\tablenotemark{a} $\times$ 10$^{-1}$   \\
\end{tabular}
\tablenotetext{a}{Fixed to the value of H$_2$CS}
\end{table}

\clearpage

\begin{table}
\caption{Relative abundances of deuterated molecules in several sources}
\vspace{0.2cm}
\begin{tabular}{l r r r r r r}
\hline
\hline
Molecule & Barnard 1 & TMC-1\tablenotemark{\it a} \rm &
TMC-1\tablenotemark{\it b} \rm & L134N & IRAS 16293-2422 & Orion (KL) \\
\hline
DCO$^+$/HCO$^+$ & & 0.012$^1$ & & 0.180$^2$ & 0.009$^3$ & $<$0.01$^4$ \\
N$_2$D$^+$/N$_2$H$^+$ & 0.150$^5$ & 0.0064$^1$ & 0.080$^2$ & 0.300$^5$ & & $<$0.30$^4$ \\
NH$_2$D/NH$_3$  & 0.128$^6$ & $<$0.001$^1$ & 0.020$^2$ & 0.10$^{2,7}$ & 0.100$^3$ & 0.062$^8$ \\
ND$_2$H/NH$_2$D & & & & 0.050$^7$ & &  \\
ND$_3$/NH$_3$ & 0.0008$^9$ & & & & &  \\
\hline
HDCO/H$_2$CO & 0.143 & & 0.059$^1$& 0.068$^1$ & 0.140$^3$ & 0.140$^8$ \\
D$_2$CO/HDCO & 0.400 & & & & $<$0.50$^7$ & 0.021$^8$ \\
HDS/H$_2$S & & & &  & 0.100$^3$ & $<$0.02$^{11}$ \\
D$_2$S/HDS & 0.108$^{12}$ & & & & $<$0.264$^{12}$ & \\
HDCS/H$_2$CS & 0.300 & 0.020$^{13}$ & & & & \\
D$_2$CS/HDCS & 0.333 & & & & & \\
\hline
CH$_3$OD/CH$_3$OH & & 0.026$^1$ & & $<$0.032$^1$ & 0.015$^{14}$ & 0.01-0.06$^{15}$ \\
CH$_2$DOH/CH$_3$OH & & & & & 0.306$^{14}$ & \\
CHD$_2$OH/CH$_3$OH & & & & & 0.060$^{14}$ & \\
CD$_3$OH/CH$_3$OH & & & & & $\sim$0.010$^{14}$ & \\
\end{tabular}
\tablenotetext{a}{Cyanopolyyne peak}
\tablenotetext{b}{Ammonia peak}
\tablerefs{
1. \cite{Tur01};
2. \cite{Tin00};
3. \cite{van95};
4. \cite{Tur78};
5. \cite{Ger01};
6. \cite{Sai00};
7. \cite{Rou00};
8. \cite{Tur90};
9. \cite{Lis02};
10. \cite{Loi00};
11. \cite{Minh90};
12. \cite{Vas03};
13. \cite{Min97};
14. \cite{Par04};
15. \cite{Mau88}.}

\end{table}

\clearpage

\begin{table}
  \caption{Fractionation reactions introduced in the network}
\vspace{0.2cm}
  \begin{tabular}{lcc}
\hline
\hline
 Reaction & k$_f$\tablenotemark{a}  &  $\beta$ \\
           & (cm$^3$ s$^{-1}$) &   (K) \\
\hline
H$_3$CO$^+$ +   HD  $\rightleftharpoons$   H$_2$DCO$^+$ + H$_2$ &                      2.0 $\times$ 10$^{-9}$ &  500  \\
H$_3$S$^+$ +  HD   $\rightleftharpoons$  H$_2$DS$^+$ +   H$_2$ &  2.0 $\times$ 10$^{-9}$ &    500 \\                  
H$_2$DS$^+$ +  HD $\rightleftharpoons$ HD$_2$S$^+$ +  H$_2$   &   2.0 $\times$ 10$^{-9}$ &    400 \\                  
HD$_2$S$^+$ + HD   $\rightleftharpoons$ D$_3$S$^+$ +  H$_2$  &  2.0 $\times$  10$^{-9}$ &    300 \\               
H$_3$S$^+$  + D$_2$   $\rightleftharpoons$  HD$_2$S$^+$ + H$_2$  & 2.0 $\times$ 10$^{-9}$ &    500 \\                  
H$_2$DS$^+$ +   D$_2$    $\rightleftharpoons$  D$_3$S$^+$ + H$_2$  & 2.0 $\times$ 10$^{-9}$ &    400 \\                   
H$_2$DS$^+$   D$_2$  $\rightarrow$  HD$_2$S$^+$ + HD & 2.0 $\times$ 10$^{-9}$ &      400 \\                     
    HD$_2$S$^+$ +   D$_2$  $\rightleftharpoons$  D$_3$S$^+$ + HD  & 2.0 $\times$ 10$^{-9}$ &    300 \\                     
\hline
H$_2$CS$^+$ +   HD $\rightleftharpoons$     HDCS$^+$   H$_2$ &    2.0 $\times$ 10$^{-9}$ &    500 \\  
HDCS$^+$ +  HD  $\rightleftharpoons$     D$_2$CS$^+$   H$_2$ &    2.0 $\times$ 10$^{-9}$ &    400 \\ 
H$_2$CS$^+$ +   D$_2$ $\rightleftharpoons$     D$_2$CS$^+$   H$_2$ &    2.0 $\times$ 10$^{-9}$ &    500 \\   
H$_2$CS$^+$ +   D$_2$ $\rightleftharpoons$     HDCS$^+$   HD &    2.0 $\times$ 10$^{-9}$ &    400 \\    
HDCS$^+$ +  D$_2$ $\rightleftharpoons$     D$_2$CS$^+$   HD &    2.0 $\times$ 10$^{-9}$ &    300 \\      
H$_3$CS$^+$ +   HD $\rightleftharpoons$     H$_2$DCS$^+$   H$_2$ &    2.0 $\times$ 10$^{-9}$ &    500 \\  
H$_2$DCS$^+$ +  HD  $\rightleftharpoons$     HD$_2$CS$^+$   H$_2$ &    2.0 $\times$ 10$^{-9}$ &    400 \\ 
HD$_2$CS$^+$ +  HD  $\rightleftharpoons$     D$_3$CS$^+$   H$_2$ &    2.0 $\times$ 10$^{-9}$ &    400 \\
H$_3$CS$^+$ +   D$_2$ $\rightleftharpoons$     HD$_2$CS$^+$   H$_2$ &    2.0 $\times$ 10$^{-9}$ &    500 \\   
H$_3$CS$^+$ +   D$_2$ $\rightleftharpoons$   H$_2$DCS$^+$   H$_2$ &    2.0 $\times$ 10$^{-9}$ &    400 \\   
H$_2$DCS$^+$ +   D$_2$ $\rightleftharpoons$     HD$_2$CS$^+$   HD &    2.0 $\times$ 10$^{-9}$ &    500 \\    
HD$_2$CS$^+$ +  D$_2$ $\rightleftharpoons$     D$_3$CS$^+$   HD &    2.0 $\times$ 10$^{-9}$ &    500 \\      
\hline
H$_3$S$^+$ +  D $\rightleftharpoons$   H$_2$DS$^+$ +  H &  2.0 $\times$ 10$^{-9}$ &    500 \\      
H$_2$DS$^+$ +  D $\rightleftharpoons$   HD$_2$S$^+$ +  H &  2.0 $\times$ 10$^{-9}$ &    400 \\      
HD$_2$S$^+$ +  D $\rightleftharpoons$   D$_3$S$^+$ +  H &  2.0 $\times$ 10$^{-9}$ &    300 \\      
\hline
H$_2$CS$^+$ +   D   $\rightleftharpoons$   HDCS$^+$ +  H &  2.0 $\times$ 10$^{-9}$ &    500 \\   
HDCS$^+$ + D $\rightleftharpoons$   D$_2$CS$^+$ +  H &  2.0 $\times$ 10$^{-9}$ &    400 \\
H$_3$CS$^+$ +   D$\rightleftharpoons$   H$_2$DCS$^+$ +  H &  2.0 $\times$ 10$^{-9}$ &    500 \\    
H$_2$DCS$^+$ +   D   $\rightleftharpoons$   HD$_2$CS$^+$ +  H &  2.0 $\times$ 10$^{-9}$ &    400 \\   
HD$_2$CS$^+$ +   D   $\rightleftharpoons$   D$_3$CS$^+$ +  H &  2.0 $\times$ 10$^{-9}$ &    300 \\   
\end{tabular}
\tablenotetext{a}{The forward rate is k$_f$; the reverse rate, k$_r$, is given by k$_f$ e$^{-\beta/T}$}
\end{table}

\clearpage

\begin{table}
  \caption{Prediction of sulfur containing molecules fractional abundances and deuterium fractionation.}
\vspace{0.2cm}
  \begin{tabular}{llll}
\hline 
\hline
\noalign{\smallskip}
 model conditions  & Model 1 & Model 2 & Model 3 \\
\hline
\noalign{\smallskip}
H$_2$ (cm$^{-3}$) & 10$^4$ & 10$^5$ & 10$^6$\\
C/H$_2$  & $7.5 \times 10^{-5}$ & $3.0 \times 10^{-5}$ & $1.5 \times 10^{-5}$ \\
O/H$_2$  & $2 \times 10^{-4}$ & $8 \times 10^{-5}$ & $4 \times 10^{-5}$ \\
N/H$_2$  & $2 \times 10^{-5}$ & $2 \times 10^{-5}$ & $2 \times 10^{-5}$ \\
representative molecular cloud & TMC-1  & Barnard 1  & L134N \\
\hline
steady state results  &  &  &  \\
\hline
H (cm$^{-3}$)  & 2.24  & 2.22   & 2.25  \\
HD (cm$^{-3}$)  & $2.76 \times 10^{-1}$ & 2.76   & 26.8 \\
\hline
x(e$^-$)\tablenotemark{\it a} \rm  & $6.03 \times 10^{-8}$ &  $3.52 \times
10^{-8}$  & $2.60 \times 10^{-8}$ \\
x(H$_2$CO)\tablenotemark{\it b} \rm & $ 1.2\times 10^{-9}$ & $ 4.7\times
10^{-10}$ & $ 1.1\times  10^{-10}$ \\
  & $ 1.3\times 10^{-9}$ & $ 5.0\times 10^{-10}$ & $ 1.1\times 10^{-10}$ \\
x(H$_2$S)\tablenotemark{\it b} \rm & $ 4.5\times 10^{-10}$ & $ 8.1\times
10^{-10}$ & $  8.7\times10^{-10}$ \\
 & $ 6.1\times 10^{-10}$ & $ 1.4\times 10^{-9}$ & $ 1.7\times 10^{-9}$ \\
x(H$_2$CS)\tablenotemark{\it b} \rm & $ 4.1\times 10^{-11}$ & $ 2.9\times
10^{-11}$ & $ 1.0\times  10^{-11}$ \\
 & $ 4.8\times 10^{-11}$ & $ 2.9\times 10^{-11}$ & $ 1.3\times 10^{-11}$ \\
\hline
D/H & 0.0053 & 0.012 & 0.021 \\
D$_2$/HD & 0.021 & 0.036 & 0.059 \\
H$_2$D$^+$/H$_3$$^+$ & 0.029 & 0.10 & 0.16 \\
D$_2$H$^+$/H$_2$D$^+$ & 0.054 & 0.10 & 0.16 \\
D$_3$$^+$/D$_2$H$^+$ & 0.041 & 0.086 & 0.15 \\
\hline
HDCO/H$_2$CO\tablenotemark{\it b} \rm & 0.091  & 0.12  & 0.14  \\
                  & 0.048  & 0.07 & 0.089 \\
D$_2$CO/HDCO\tablenotemark{\it b} \rm & 0.34  & 0.57  & 0.80  \\
                  & 0.60  & 0.95   & 1.20   \\
HDS/H$_2$S\tablenotemark{\it b} \rm  & 0.26   & 0.37  & 0.43\\
               & 0.017  & 0.032  & 0.050  \\
D$_2$S/HDS\tablenotemark{\it b} \rm & 0.53   & 1.0 & 1.32 \\
                 & 0.0049  & 0.007  & 0.0086  \\
HDCS/H$_2$CS\tablenotemark{\it b} \rm & 0.16  & 0.18  & 0.18  \\
                  & 0.041   & 0.065 & 0.086  \\
D$_2$CS/HDCS\tablenotemark{\it b} \rm & 0.48 & 0.85 & 1.17  \\
                  & 0.37  & 0.67   & 0.92  \\
\hline
DCO$^+$/HCO$^+$ & 0.028  & 0.043  & 0.066   \\
NH$_2$D/NH$_3$ & 0.040    & 0.077   & 0.20  \\
ND$_2$H/NH$_2$D & 0.027  & 0.041     & 0.051 \\
ND$_3$/ND$_2$H & 0.035  & 0.05    & 0.056  \\
N$_2$D$^+$/N$_2$H$^+$ & 0.032  & 0.047  & 0.068  \\

\end{tabular}
\tablenotetext{a}{The x symbol refers to the fractional abundance relative to
molecular  hydrogen H$_2$.}
\tablenotetext{b}{The first line and second lines correspond respectively to
models   with and without the  fractionation reactions of H$_2$CO$^+$,
H$_3$CO$^+$, H$_3$S$^+$, H$_3$CS$^+$}
\end{table}

\end{document}